\documentclass[12pt]{article}
\usepackage{times}
\usepackage{epsf}
\usepackage{graphicx}

\def\beq{\begin{equation}}
\def\eeq{\end{equation}}
\def\bea{\begin{eqnarray}}
\def\eea{\end{eqnarray}}
\def\ba{\begin{array}}
\def\ea{\end{array}}
\def\bea{\begin{eqnarray}}
\def\eea{\end{eqnarray}}

\begin{document}
\title{Exponential Type
Complex and non-Hermitian Potentials within Quantum Hamilton-Jacobi
Formalism}
\author{\vspace{1cm}
         {\"Ozlem Ye\c{s}ilta\c{s} $^1$}
         \,\, and \,\,
        { Ramazan Sever $^2$ $^,$}
         \thanks{Corresponding author: sever@metu.edu.tr}
\\
{\small \sl $^1$ Turkish Atomic Energy Authority, Nuclear Fusion and
Plasma Physics Laboratory, 06983, Ankara, Turkey}
\\{\small \sl $^2$ Department
of Physics, Middle East Technical University, 06531, Ankara, Turkey
}}
\maketitle
\begin{abstract}
\noindent PT-/non-PT-symmetric   and   non-Hermitian   deformed
Morse and P\"{o}schl-Teller    potentials    are    studied first
time by quantum Hamilton-Jacobi approach.  Energy  eigenvalues and
eigenfunctions are   obtained   by   solving quantum Hamilton-Jacobi
equation.
\end{abstract}
~~~~~~~~~~{\small \sl PACS Ref: 03.65.Db, 03.65.Ge}


\newpage
\section{Introduction}

\indent Recently, the so called PT-symmetric quantum mechanics has
attracted wide attention [1]. In Bender and Boettcher's work in 1998
[1], it was shown that the class of non-Hermitian, Hamilton
operators such as $H =p^{2}+x^{2}(ix)^{\epsilon} ( \epsilon
> 0)$ has a real spectrum due to its PT-symmetry where P and T are the parity and
time reversal operators respectively [2]. Exact solution of the
Schr\"{o}dinger equation for various potentials which are complex
are generally of interest. It is also known that PT-symmetry does
not necessarily lead to completely real spectrum, and an extensive
kind of potentials of real or complex form are being faced with in
various fields of physics. In particular, the spectrum of the
Hamiltonian is real if PT-symmetry is not spontaneously broken.
Recently, Mostafazadeh has generalized PT symmetry by
pseudo-Hermiticity  [3]. In fact, a Hamiltonian of this type is said
to be $\eta$- pseudo Hermitian if $H^{+}=\eta H \eta^{-1}$, where
$+$ denotes the operator of adjoint. In [4] new class of
non-Hermitian Hamiltonians with real spectra was proposed which are
obtained using pseudo-symmetry. Moreover, completeness and
orthonormality conditions for eigenstates of such potentials are
proposed [5]. In the study of PT-invariant potentials various
techniques from a great variety of quantum mechanical fields have
been applied such as variational methods, numerical approaches,
Fourier analysis, semi-classical estimates, quantum field theory and
Lie group theoretical approaches [5-14]. In addition, PT-symmetric
and non-PT symmetric and also non-Hermitian potential cases such as
oscillator type potentials [15], a variety of potentials within the
framework of SUSYQM [16-19], exponential type screened potentials
[20], quasi/conditionally exactly solvable ones [21], PT-symmetric
and non-PT symmetric and also non-Hermitian potential cases within
the framework of SUSYQM via Hamiltonian Hierarchy Method [22] and
some others are studied [23-25].

\noindent The QHJ formalism, which is a formulation of quantum
mechanics was investigated as a theory related to the classical
transformation theory [26-27]. It was formulated by Leacock and
Padgett in 1983 [28-29]. Within the Quantum Hamiltonian Jacobi
approach (QHJ), which follows classical mechanics, not only the the
energy spectrum of exactly solvable (ES) and quasi-exactly solvable
(QES) models in quantum mechanics but eigenfunctions can also be
determined [30-36]. The advantage of this method is that it is
possible to determine the energy eigenvalues without having to solve
for the eigenfunctions. In this formalism, singularity structure of
the quantum momentum function $p(x)$ which is a quantum analog of
classical momentum function $p_{c}$ determines the eigenvalues of
the Hamiltonian. An exact quantization condition is formulated as a
contour integral, representing the quantum action variable, in the
complex plane. The quantization condition leads to the number nodes
of the wave function. The wavefunction is related to the quantum
momentum function (QMF). The equation satisfied by the QMF is a
non-linear differential equation, called as quantum Hamilton-Jacobi
equation. There is a boundary condition in the limit QMF which is
used to determine physically acceptable solutions for the QMF
[29-37]. In the applications, Ranjani and her collaborators applied
the QHJ formalism, to Hamiltonians with Khare-Mandal potential and
Scarf potential, characterized by discrete parity and time reversal
(PT) symmetries [31].

\noindent The purpose of the present work has been to apply the QHJ
formalism to the Hamiltonian in one dimension with non-hermitian
exponential type potentials in order to see possible singularities
for the QMF which determine eigenvalues and convenient
eigenfunctions.

\noindent The organization of the paper is as follows. In Sec. II, we briefly
introduce the Quantum Hamilton-Jacobi formalism. In Sec. III and IV,solutions of
PT-/non-PT-symmetric and non-Hermitian forms of the well-known  potentials are
presented by using QHJ method. We discuss the results in Sec. V.

\section{Quantum Hamilton-Jacobi Formalism}
\noindent In quantum theory, one assumes that function $W(x,E)$
satisfies $(2m=1)$ [33],
\begin{eqnarray}
-i\hbar \frac{\partial^{2}W(x,E)}{\partial x^{2}}+\left[\frac{\partial
W(x,E)}{\partial x}\right]^{2}=(E-V(x))
\end{eqnarray}

\noindent Eq. (1) will be called as the QHJ equation. The momentum function

\begin{eqnarray}
p(x,E)=\frac{\partial W(x,E)}{\partial x}
\end{eqnarray}

\noindent will be called as the QMF. In the limit $\hbar\rightarrow 0$, the QHJ
equation turns into the classical Hamilton-Jacobi equation. Then, QMF turns into the
classical momentum function in the $\hbar\rightarrow0$ limit:

\begin{eqnarray}
p(x,E)\rightarrow p_{c}(x,E)=\sqrt{E-V(x)}
\end{eqnarray}

\noindent In terms of $p(x,E)$ the QHJ equation, Eq.(1) can be
written as

\begin{eqnarray}
p^{2}(x,E)-i\hbar p^{'}(x,E)-[E-V(x)]=0.
\end{eqnarray}

\noindent Leacock and Padgett [28,29] proposed using the following
quantization condition for the bound states in order to obtain
eigenvalues. $C$ is a contour that encloses the moving poles between
the classical turning points and the integral

\begin{eqnarray}
J(E)=\frac{1}{2\pi}\oint_{C} p(x)\ dx
\end{eqnarray}

\noindent is called the quantum action variable. More details can be
found in the paper of Bhalla et all [30-37]. Then

\begin{eqnarray}
J=n\hbar=J(E)
\end{eqnarray}

\noindent gives the exact energy eigenvalues ($n=0,1,2,...$)
[28-33]. Leacock [28,29] defines the wave function in order to
connect QHJ equation to the Sch\"{o}dinger equation,

\begin{eqnarray}
\psi(x,E)\equiv exp\left[\frac{i}{\hbar}W(x,E)\right]
\end{eqnarray}

\noindent hence $\psi(x,E)$ satisfies the Sch\"{o}dinger equation and the physical
boundary conditions. The quantization condition becomes [33]

\begin{eqnarray}
\oint_{C} p(x,E) dx=2\pi i \sum_{k} (Res)_{k}=nh
\end{eqnarray}

\noindent where $\sum_{k} (Res)_{k}$ is sum of the residues. In the
QHJ equation, if $V(x)$ has a singular point, $p(x,E)$ will also
have singular point in that zone [33]. These singularities are known
as fixed singular points which are energy independent. Other types
of singular points are the moving singular points. They can only be
poles with residue $-i\hbar$. Suppose $b\ne0$ then, moving
singularities are in the form of [30-37]

\begin{eqnarray}
p(x,E)\sim \frac{b}{(x-x_{0})^{r}}+...
\end{eqnarray}

\noindent in the QHJ equation. If the potential is not singular at
$x=x_{0}$ then $r$ must be equal to one and $b=-i\hbar$ [33].

\section{Generalized Morse Potential}

\noindent The generalized Morse potential is given by [19]

\begin{eqnarray}
V(x)=V_{1}e^{-2\alpha x}-V_{2}e^{-\alpha x}
\end{eqnarray}

\noindent In order to apply QHJ method, we write the potential
relation in Eq.(4) $(\hbar=2m=1)$

\begin{eqnarray}
p^{2}-ip^{'}-[E-V_{1}e^{-2\alpha x}+V_{2}e^{-\alpha x}]=0
\end{eqnarray}

\noindent Substitution of the transformation  of
$y=\sqrt{V_{1}}e^{-\alpha x}$ in Eq.(11) gives:

\begin{eqnarray}
p^{2}(y,E)+i \alpha y p^{'}(y,E)-\left[E-y^{2}+\frac{V_{2}}{\sqrt{V_{1}}}y\right]=0
\end{eqnarray}

\noindent Define $p=i \alpha y \phi$ and $\chi=\phi+\frac{1}{2y}$ in
order to transform Eq.(11) into a Riccati type differential equation
as,

\begin{eqnarray}
\chi^{'}+\chi^{2}+\frac{1}{4y^{2}}+\frac{1}{\alpha^{2}
y^{2}}\left[E-y^{2}+\frac{V_{2}}{\sqrt{V_{1}}}y\right]=0
\end{eqnarray}

\noindent As it can be seen from Eq. (13), $\chi$ has a pole only at
$y=0$ and for $y=0$ define $\chi$ as,

\begin{eqnarray}
\chi=\frac{b_{1}}{y}+a_{0}+a_{1}y
\end{eqnarray}

\noindent Substitute Eq.(14) in (13) and equate coefficients of
$\frac{1}{y^{2}}$ yields

\begin{eqnarray}
b_{1}=\frac{1}{2\alpha} (\alpha\pm2\sqrt{-E})
\end{eqnarray}

\noindent When it comes to the discussion of the behaviour of $\chi$
at infinity, one expands $\chi$ as:

\begin{eqnarray}
\chi=a_{0}+\frac{\lambda}{y}+\frac{\lambda_{1}}{y^{2}}
\end{eqnarray}

\noindent and find $\lambda$ as

\begin{eqnarray}
\lambda=\pm \frac{V_{2}}{2\alpha \sqrt{V_{1}}}
\end{eqnarray}

\noindent One can see that the behaviour of $\chi$ is
$\frac{b1+n}{y}$ for large $y$. Hence,

\begin{eqnarray}
b_{1}+n=\lambda
\end{eqnarray}

\noindent In order to find the wavefunction, $\chi(y)$ can be
written as the sum of the Laurent expansions around different
singular points, plus a constant $C_{1}$. Hence

\begin{eqnarray}
\chi(y)=\frac{b_{1}}{y}+\frac{P_{n}^{'}(y)}{P_{n}(y)}+C_{1}
\end{eqnarray}

\noindent where $P_{n}(y)$ is a $n$ th degree polynomial. Substitute
Eq. (19) in (13) and get

\begin{eqnarray}
\frac{P_{n}^{''}}{P_{n}}+\frac{2P_{n}^{'}}{P_{n}}\left(\frac{2
b_{1}}{y}+
2C_{1}\right)+\left(\frac{b^{2}_{1}-b_{1}+E/\alpha^{2}+1/4}{y^{2}}+
\frac{2b_{1}C_{1}}{y}+C^{2}_{1}-\frac{1}{\alpha^{2}}+\frac{V_{2}}
{\alpha^{2}y\sqrt{V_{1}}}\right)=0.
\end{eqnarray}

\noindent For large $y$ one can find $C_{1}=\pm \frac{1}{\alpha}$.
The wave function in terms of $\chi$ can be written by using eq.
(19) and (7) as

\begin{eqnarray}
\psi(y)=exp\left(\int\left(\frac{b_{1}}{y}+\frac{P_{n}^{'}}{P_{n}}-\frac{1}{\alpha}-
\frac{1}{2y}\right)dy\right).
\end{eqnarray}

\noindent In Eq. (21) the correct value of $C_{1}$ is used as
$C_{1}=- \frac{1}{\alpha}$ because of the condition for the
wavefunction which is known as $y\rightarrow \infty$,
$\psi(y)\rightarrow 0$. It is seen from Eq. (15) that $b_{1}$ has
two values and no particular value has been chosen. Using Eq. (18),
the energy eigenvalues for any n-th state become,

\begin{eqnarray}
E_{n}=-\frac{\alpha^{2}}{4}\left[-(2n+1)+\frac{V_{2}}{\alpha
\sqrt{V_{1}}}\right]^{2}
\end{eqnarray}

\noindent If we use Eq.(22),(15) and $C_{1}=-1/\alpha$ in Eq.(20),
it becomes

\begin{eqnarray}
y P_{n}^{''}(y)+(\frac{V_{2}}{2\alpha
\sqrt{V_{1}}}+1-y)P_{n}^{'}(y)+n P_{n}(y)=0
\end{eqnarray}

\noindent which is a Laguerre differential equation. Therefore, the
wavefunction is obtained as

\begin{eqnarray}
\psi_{n}(y)=C_{n}\,\ e^{-\frac{y}{\alpha}}\,\ y^{(-n\pm
\frac{V_{2}}{2\alpha \sqrt{V_{1}}})}\,\ L^{\frac{V_{2}}{2\alpha
\sqrt{V_{1}}}}_{n}(y).
\end{eqnarray}

\noindent where $C_{n}$ is a normalization constant and
$L^{\frac{V_{2}}{2\alpha \sqrt{V_{1}}}}_{n}(y)$ are Laguerre
functions. The wavefunction satisfies the boundary condition that is
$y \rightarrow \infty$, \,\ $\psi(y) \rightarrow 0$.

\subsection{Non-PT symmetric and non-Hermitian Morse Potential}

\noindent In  equation (10), if the potential parameters are defined
as $V_{1}=(A+iB)^{2}$, $V_{2}=(2C+1)(A+iB)$ and $\alpha=1$, then the
potential becomes [19],

\begin{eqnarray}
V(x)=(A+iB)^{2}e^{-2x}-(2C+1)(A+iB)e^{-x}
\end{eqnarray}

\noindent where $A$, $B$ and $C$ are arbitrary real parameters and
$i=\sqrt{-1}$. The QHJ equation is

\begin{eqnarray}
p^{2}-ip^{'}-[E-(A+iB)^{2} e^{-2x}+(2C+1)(A+iB)e^{-x}]=0
\end{eqnarray}

\noindent Using the transformation in the form of $y=(A+iB)e^{-x}$
in the Eq. (26), then using $p(y)=iy\phi$ and
$\chi=\phi+\frac{1}{2y}$, Eq. (26) becomes

\begin{eqnarray}
\chi^{'}+\chi^{2}+\frac{1}{4y^{2}}+\frac{1}{y^{2}}\left[E-y^{2}+(2C+1)y\right]=0
\end{eqnarray}

\noindent As it is seen from Eq. (27), $\chi$ has a pole only at
$y=0$ and for $y=0$ define $\chi$ for the Eq.(27) as,

\begin{eqnarray}
\chi=\frac{b_{1}}{y}+a_{0}+a_{1}y
\end{eqnarray}

\noindent Using the Eq. (28) in (27), $b_{1}$ is found as

\begin{eqnarray}
b_{1}=\frac{1}{2}\pm i\sqrt{\frac{|4E|}{2}}
\end{eqnarray}

\noindent Following the same procedure that is given in section 3,
energy is found as

\begin{eqnarray}
E_{n}=-(n-C)^{2}.
\end{eqnarray}

\noindent Now choose the potential parameters in Eq.(10) as $V_{1}$
is real and $V_{2}=A+iB$, the Morse potential can be written in the
following form

\begin{eqnarray}
V(x)=V_{1}e^{-2i\alpha x}-(A+iB)e^{-i\alpha x}
\end{eqnarray}

\noindent and following the same procedure, the energy is obtained as

\begin{eqnarray}
E_{n}=\alpha^{4} \left[(n+1/2)-\frac{A+iB}{2\alpha \sqrt{|-V_{1}|}}\right]^{2}
\end{eqnarray}

\noindent According to Eq. (32), the spectrum is real in the case of
$Im(V_{2})=0$.
\\

\subsection{PT symmetric and non-Hermitian Morse Potential}

\noindent When $\alpha=i\alpha$ and $V_{1}, V_{2}$ are real, the Morse potential
becomes

\begin{eqnarray}
V(x)=V_{1}e^{-2i \alpha x}-V_{2}e^{-i\alpha x}
\end{eqnarray}

\noindent Thus, the energy eigenvalues are obtained as

\begin{eqnarray}
E_{n}=\alpha^{4}\left[(n+\frac{1}{2})+\frac{V_{2}}{2\alpha
\sqrt{|-V_{1}|}}\right]^{2}
\end{eqnarray}

\noindent If we take the parameters of Eq.(39) as
$V_{1}=-\omega^{2}$, $V_{2}=D$ and $\alpha=2$ then, corresponding
eigenvalues for any n-th state are obtained as

\begin{eqnarray}
E_{n}=(2n+1+\frac{D}{2\omega})^{2}
\end{eqnarray}

\noindent are consistent with the results [10-11,19].

\section{P\"{o}schl-Teller Potential}

\noindent The general form of the P\"{o}schl-Teller potential is

\begin{eqnarray}
V(x)=-4V_{0}\frac{e^{-2\alpha x}}{(1+qe^{-2\alpha x})^{2}}
\end{eqnarray}

\noindent The QHJ equation is given as

\begin{eqnarray}
p^{2}-i p^{'}-\left(E+4V_{0}\frac{e^{-2\alpha x}}{(1+qe^{-2\alpha x})^{2}}\right)=0
\end{eqnarray}

\noindent If we take $y=\pm i\sqrt{q}e^{-\alpha x}$ and use the transformations as
$p=-i\alpha y\phi$ and $\chi=\phi+\frac{1}{2y}$, the QHJ equation turns into

\begin{eqnarray}
\chi^{'}+\chi^{2}+\frac{1}{4y^{2}}+\frac{1}{\alpha^{2}y^{2}}\left[E-\frac{4V_{0}}{q}
\frac{y^{2}}{(1-y^{2})^{2}}\right]=0.
\end{eqnarray}

\noindent As it is seen from Eq.(38), $\chi$ has poles at $y=0$ and
$\pm 1$. $\chi$ is expanded at $y=0$ and $b_{1}$ is found

\begin{eqnarray}
b_{1}=\frac{1}{2\alpha}(\alpha\pm2\sqrt{-E})
\end{eqnarray}

\noindent At $y=1$, one can expand $\chi$ as

\begin{eqnarray}
\chi=\frac{b^{'}_{1}}{1-y}+a^{'}_{0}+a^{'}_{1}(1-y)
\end{eqnarray}

\noindent and $b^{'}_{1}$ is found as

\begin{eqnarray}
b^{'}_{1}=\frac{1}{2q\alpha}(q\alpha\pm\sqrt{\alpha^{2}q^{2}+8qV_{0}})
\end{eqnarray}

\noindent At $y=-1$, one can expand $\chi$ as

\begin{eqnarray}
\chi=\frac{b^{''}_{1}}{1+y}+a^{''}_{0}+a^{''}_{1}(1+y)
\end{eqnarray}

\noindent Substitute Eq. (42) in (38), to obtain
$b^{'}_{1}=b^{''}_{1}$. One can look at the behavior of $\chi$ at
infinity with expanding $\chi$ as

\begin{eqnarray}
\chi=A_{0}+\frac{\lambda}{y}+\frac{\lambda}{y^{2}}
\end{eqnarray}

\noindent From  Eq. (43) and (38), $\lambda$ is found as

\begin{eqnarray}
\lambda=\frac{1}{2\alpha}(\alpha\pm2\sqrt{-E})
\end{eqnarray}

\noindent and behavior of $\chi$ is
$\frac{b1+b^{'}_{1}+b^{''}_{1}+2n}{y}$ for large $y$. Hence,

\begin{eqnarray}
\lambda=b_{1}+b^{'}_{1}+b^{''}_{1}+2n
\end{eqnarray}

\noindent In order to find the wavefunctions, $\chi$ is written as

\begin{eqnarray}
\chi=\frac{b_{1}}{y}+\frac{b^{'}_{1}}{1-y}+\frac{b^{''}_{1}}{1+y}+\frac{P_{n}^{'}(y)}{P_{n}(y)}+C_{2}
\end{eqnarray}

\noindent Substituting Eq. (46) in (38), $C_{2}$  can be found as
$C_{2}=0$ for large $y$. The wavefunction can be written as

\begin{eqnarray}
\psi=exp\left(\int\left(\frac{2b_{1}-1}{2y}+\frac{b^{'}_{1}}{1-y}+\frac{b^{''}_{1}}{1+y}+
\frac{P_{n}^{'}(y)}{P_{n}(y)}\right)dy\right)
\end{eqnarray}

\noindent If we look at Eq. (47), $b_{1}$  and $b^{'}_{1}$ have two
values and both of them gives appropriate results for energy
spectrum and wavefunction, no particular value has been chosen.
Thus, residues are given as $b_{1}=\frac{1}{2\alpha}(\alpha\pm
2\sqrt{-E})$ and $b^{'}_{1}=b^{''}_{1}=\frac{1}{2q\alpha}(q \alpha
\pm \sqrt{\alpha^{2}q^{2}+8qV_{0}})$. Finally the energy is obtained
by using Eq. (45) as

\begin{eqnarray}
E_{n}=-\frac{\alpha^{2}}{4} \left((2n+1)\pm
\sqrt{1+\frac{8V_{0}}{q\alpha^{2}}}\right)^{2}
\end{eqnarray}

\noindent Using Eqs. (47,48) and (38), one can find the wave
function as

\begin{eqnarray}
\psi_{n}(y)=N\,\ y^{-(n-1/2)\pm \gamma}(1-y^{2})^{\frac{1}{2}(1\pm
\gamma)}\,\ P^{-\nu_{2}-\frac{1}{2},\,\ \nu_{2}-\frac{1}{2}}_{n}(y)
\end{eqnarray}

\noindent where $N$ is a normalization constant,
$\gamma=\sqrt{1+\frac{8V_{0}}{q\alpha^{2}}}$,\,\
$P^{-\nu_{2}-\frac{1}{2},\nu_{2}-\frac{1}{2}}_{n}(y)$ stands for
Jacobi polynomials and $\nu_{2}=\sqrt{\frac{8V_{0}}{q\alpha^{2}}}$.
If we look at Eq.(49), there are three cases for physical solutions
because of the wavefunction that satisfies the boundary condition as
$y\rightarrow \infty$, \,\ $\psi(y)\rightarrow 0$. If \,\
$-(n-1/2)\pm \gamma < 0$ and $1\pm \gamma > 0$, it should be $1\pm
\gamma > |-(n-1/2)\pm \gamma|$. If \,\ $-(n-1/2)\pm \gamma < 0$ and
$1\pm \gamma < 0$, there is no restriction for the parameters and
there are physical solutions in this case. The last case can be
defined as; if \,\ $-(n-1/2)\pm \gamma > 0$, \,\ $1\pm \gamma >
|-(n-1/2)\pm \gamma|$  for appropriate solutions.

\ \

\subsection{Non-PT symmetric and non-Hermitian P\"{o}schl-Teller
cases}

\noindent Here, $V_{0}$ and $q$ are complex parameters $V_{0}=V_{0R}+iV_{0I}$ and
$q=q_{R}+iq_{I}$ but $\alpha$ is a real parameter. Although the potential is complex
and the corresponding Hamiltonian is non-Hermitian and also non-PT symmetric, there
may be real spectra if and only if $V_{0I}q_{R}=V_{0R}q_{I}$. When both parameters
$V_{0}$ and $q$ are taken pure imaginary, the potential turns out to be,

\begin{eqnarray}
V(x)=-4V_{0}\frac{2qe^{-4\alpha x}+i(1-q^{2}e^{-4\alpha x})}{(1+q^{2}e^{-4\alpha
x})^{2}}
\end{eqnarray}

\noindent For simplicity, we use the notation $V_{0}$ and $q$
instead of $V_{0I}$ and $q_{I}$. In this case, we get the same
energy eigenvalues as in Eq.(48). If $q$ is an arbitrary real
parameter and $V_{0}\Rightarrow iV_{0}$ also $\alpha\Rightarrow
i\alpha$ completely imaginary, the potential becomes

\begin{eqnarray}
V(x)=-4V_{0}\frac{(1-q^{2})sin2\alpha x+i(2q+(1+q^{2})cos2\alpha
x)}{(1+q^{2})^{2}+4qcos2\alpha x(1+qcos2\alpha x+q^{2})}
\end{eqnarray}

\noindent and the corresponding energy eigenvalues become

\begin{eqnarray}
E=\frac{\alpha^{2}}{4}\left[2n+1+\sqrt{1+\frac{8V_{0}}{\alpha^{2}}}\right]^{2}
\end{eqnarray}

\noindent For simplicity, let us take all three parameters $\alpha,
q, V_{0}$ purely imaginary. Then the potential takes the form

\begin{eqnarray}
V(x)=-4V_{0}\frac{(1+q^{2})sin2\alpha x+2q+i((1-q^{2})cos2\alpha
x)}{(1+q^{2})^{2}+4q^{2}(1-cos^{2}2\alpha x)+4q(1+q^{2})sin2\alpha x}
\end{eqnarray}

\noindent and the energy becomes

\begin{eqnarray}
E_{n}=\frac{\alpha^{2}}{4} \left[2n+1+\frac{1}{2\alpha
q}\sqrt{\alpha^{2}q^{2}+(1+q^{2})V_{0}}\right]^{2}
\end{eqnarray}

\ \

\subsection{PT symmetric and non-Hermitian  P\"{o}schl-Teller
cases}

\noindent We choose parameters $V_{0}$ and $q$ real and also
$\alpha=i \alpha$. Then, the potential turns into

\begin{eqnarray}
V(x)=-4V_{0}\frac{(1+q^{2})cos2\alpha x+2q+i(q^{2}-1)sin2\alpha x}{(1+q^{2})^{2}+4q
cos2\alpha x((1+q cos2\alpha x+q^{2})}
\end{eqnarray}

\noindent and corresponding energy spectrum is

\begin{eqnarray}
E=-\frac{\alpha^{2}}{4}\left[2n+1+\sqrt{1+\frac{8V_{0}}{\alpha^{2}}}\right]^{2}
\end{eqnarray}

\section{Conclusions}

\noindent We have applied the PT-symmetric formulation to solve the
Quantum Hamilton-Jacobi equation for Morse and P\"{o}schl-Teller
potentials in both real and complex forms. We have obtained the
energy eigenvalues and the corresponding wave functions for
different forms of these potentials within Quantum Hamilton-Jacobi
formalism. The real energy spectra of the PT-/non-PT- symmetric
complex valued non-Hermitian potentials have been obtained in case
the potential parameters are restricted. It is also shown that the
QHJ formalism is a good approach to obtain eigenfunctions and energy
eigenvalues for a class of exponential type potentials discussed
here within PT symmetric frame. As a result, we have pointed out
that our exact results of complexified general Morse and
P\"{o}schl-Teller potentials may increase the number of applications
of complex Hamiltonians with real energies in the extensive study of
different quantum systems within the flexible Quantum
Hamilton-Jacobi approach.  Finally we should state that this work is
the first application on the study of PT-symmetry for the Quantum
Hamilton-Jacobi approach.

\end{document}